\pdfoutput=1
\documentclass[aps,twocolumn,pre,8pt,showpacs,floatfix,nofootinbib,superscriptaddress
]{revtex4-1}
\usepackage{subfigure}
\usepackage{amssymb}
\usepackage{amsfonts}
\usepackage{amsmath}
\usepackage{amsthm}
\usepackage{epsfig}
\usepackage{graphicx}
\usepackage[usenames,dvipsnames]{color}
\usepackage[latin1]{inputenc}
\usepackage{comment}

\usepackage[hidelinks,unicode=true]{hyperref}
\hypersetup{colorlinks=true,
 	linkcolor=blue,
 	urlcolor=blue,
 	citecolor=blue,
 	pdfhighlight=/N
 }
 \usepackage{comment}
 \usepackage[normalem]{ulem}

\begin{document}

\title{On the quantum mechanics of how an ideal carbon nanotube field emitter can exhibit a constant field enhancement factor}

\author{Caio P. de Castro}
\email{caiobdo.fis@gmail.com}
\address{Instituto de F\'{\i}sica, Universidade Federal da Bahia,
   Campus Universit\'{a}rio da Federa\c c\~ao,
   Rua Bar\~{a}o de Jeremoabo s/n,
40170-115, Salvador, BA, Brazil}

\author{Thiago A. de Assis}
\email{thiagoaa@ufba.br}
\address{Instituto de F\'{\i}sica, Universidade Federal da Bahia,
   Campus Universit\'{a}rio da Federa\c c\~ao,
   Rua Bar\~{a}o de Jeremoabo s/n,
40170-115, Salvador, BA, Brazil}

\author{Roberto Rivelino}
\email{rivelino@ufba.br}
\address{Instituto de F\'{\i}sica, Universidade Federal da Bahia,
   Campus Universit\'{a}rio da Federa\c c\~ao,
   Rua Bar\~{a}o de Jeremoabo s/n,
40170-115, Salvador, BA, Brazil}

\author{Fernando de B. Mota}
\email{fbmota@ufba.br}
\address{Instituto de F\'{\i}sica, Universidade Federal da Bahia,
   Campus Universit\'{a}rio da Federa\c c\~ao,
   Rua Bar\~{a}o de Jeremoabo s/n,
40170-115, Salvador, BA, Brazil}

\author{Caio M. C. de Castilho}
\email{caio@ufba.br}
\address{Instituto de F\'{\i}sica, Universidade Federal da Bahia,
   Campus Universit\'{a}rio da Federa\c c\~ao,
   Rua Bar\~{a}o de Jeremoabo s/n,
40170-115, Salvador, BA, Brazil}
\address{Centro Interdisciplinar em Energia e Ambiente, Universidade Federal da Bahia, Campus Universit\'{a}rio da Federa\c{c}\~{a}o, 40170-115 Salvador, BA, Brazil}

\author{Richard G. Forbes}
\email{r.forbes@surrey.ac.uk}
\address{Advanced Technology Institute \& Department of Electrical and Electronic Engineering, University of Surrey, Guildford, Surrey
GU2 7XH, U.K.}

\begin{abstract}
Measurements of current-voltage characteristics from ideal carbon nanotube (CNT) field electron emitters of small apex radius have shown that these emitters can exhibit a linear Fowler-Nordheim (FN) plot [e.g., Dean and Chalamala, Appl. Phys. Lett., 76, 375, 2000]. From such a plot, a constant (voltage-independent) characteristic field enhancement factor (FEF) can be deduced. Over fifteen years later, this experimental result has not yet been convincingly retrieved from first-principles electronic structure calculations, or more generally from quantum mechanics (QM). On the contrary, several QM calculations have deduced that the characteristic FEF should be a function of the macroscopic field applied to the CNT. This apparent contradiction between experiment and QM theory has been an unexplained feature of CNT emission science, and has raised doubts about the ability of existing QM models to satisfactorily describe experimental CNT emission behavior. In this work we demonstrate, by means of a density functional theory analysis of single-walled CNTs ``floating" in an applied macroscopic field, the following significant result. This is that agreement between experiment, classical-conductor CNT models and QM calculations can be achieved if the latter are used to calculate (from the ``real" total-charge-density distributions initially obtained) the distributions of $\textit{induced}$ charge-density, induced local fields and induced local FEFs. The present work confirms, more reliably and in significantly greater detail than in earlier work on a different system, that this finding applies to the common ``post-on-a-conducing plane" situation of CNT field electron emission. This finding also brings out various further theoretical questions that need to be explored.
\end{abstract}

\maketitle

Carbon nanotubes (CNTs) are effective in field electron emission (FE) applications \cite{Saito10,ColeRev15,Vicent,Pascale,Nanotech2016} because, in the presence of an applied macroscopic field $F_\text{M}$, a sharp nanostructure develops high local fields $F_{\ell}(\textbf{\textit{r}})$ at points in space near its apex. At any point $\textbf{\textit{r}}$, a local field enhancement factor (FEF) can be defined by $\gamma
(\textbf{\textit{r}}) = F_{\ell}(\textbf{\textit{r}})/F_\text{M}$. In what follows, we consider the common situation where the field $F_\text{M}$ is parallel to the axis of a quasi-cylindrical CNT. The CNT itself is a ``floating CNT" capped at both ends, or (what is electrostatically equivalent) a ``half-CNT" standing upright on one of a pair of well-separated parallel plates and capped at the top end.

Applying the macroscopic field $F_\text{M}$ to a CNT changes its ``real" charge-density distribution $\rho_{\text{r}}(\textbf{\textit{r}}, F_\text{M})$ from that existing in the case $F_\text{M}=0$. The change is mainly in the electron density distribution, although there may also be small movements in the positions of the carbon-atom nuclei. The related ``real-local-field" distribution can be used to define a distribution of ``real" FEFs $\gamma_\text{r}(\textbf{\textit{r}})$.

It also possible to determine an ``induced-charge-density" distribution $\rho_{\text{i}}$ from the formula
\begin{equation}
\rho_{\text{i}}(\textbf{\textit{r}}, F_\text{M})\equiv \rho_{\text{r}}(\textbf{\textit{r}}, F_\text{M}) - \rho_{\text{r}}(\textbf{\textit{r}}, 0).
\label{indcharge}
\end{equation}
The resulting ``induced-local-field" distribution can be used to define a distribution of ``induced" FEFs $\gamma_\text{i}(\textbf{\textit{r}})$.

The induced-charge-density distribution can also be regarded as formally derived from the zero-$F_\text{M}$ real-charge-density distribution by means of an appropriately-defined linear response function \cite{parr1994density}. The response function will be different for each system, and will be a functional of the ground-state charge distribution.

In general terms, a FE device/system is described as $\textit{ideal}$ if the fields that determine the emission process are (at each relevant location) proportional to the measured voltage $V_\text{m}$. An ideal FE device/system is further described as $\textit{orthodox}$ if (for the purposes of modeling and testing) it is a sufficiently good approximation to assume that emission occurs by tunneling through a Schottky-Nordheim (SN) barrier.

An orthodox FE device/system will generate an effectively linear FN plot. This can be used to deduce emitter characterization parameters--including, in the system geometries of interest in this paper--a field enhancement factor. There exists an ``orthodoxy test" \cite{Forbes2013} that can be applied to a FN plot, in order to establish whether the emission is orthodox. If the test is failed, then a FEF value deduced from the plot is likely to be spurious. If an FE device/system in non-ideal, due to the existence of significant ``complications", such as (amongst others) series resistance in the current path to the high-voltage generator, or current dependence in field enhancement factors, then it is expected to fail the orthodoxy test.

Although the orthodoxy test was originally derived on the basis of the SN barrier (which is a ``planar image-rounded" barrier), the test itself is formulated in such a way that it is adequately applicable if the barrier is a rounded quasi-triangular barrier of form slightly or somewhat different from the SN barrier. Thus, the test is considered (Forbes, unpublished work) to operate adequately for FE from emitters (including CNTs) that have a rounded emitter apex and hence a tunnelling barrier slightly or somewhat different from the SN barrier. (The theoretical point is that small ``errors" in the slope correction function ``s" lead only to small ``errors" in the extraction of values of the scaled (barrier) field ``$f$", and the ``triage" nature of the test appears to cope adequately with this situation.)

In experimental CNT literature, there exist papers (e.g., \cite{Dean2000,BonardPRL2002,Xu2005}) that report linear FN plots that can be shown to pass the orthodoxy test. In the context of the usual formulations of FE theory, as applied to FE data interpretation \cite{ForbesJordan}, this suggests that--for the CNTs concerned--there is an empirical ``characteristic field enhancement factor" $\gamma_{\text{C}}$ that is a $\textit{constant}$, independent of the macroscopic field $F_\text{M}$.
This, of course, is what is automatically present in classical-conductor ``post-like" models, where the characteristic FEF is normally identified with the FEF at the apex of the classical-conductor model.

However, in previous theoretical work on FE from CNTs, characteristic-FEF values have been deduced that are not constant but depend on the value of $F_\text{M}$. For example, Zheng \textit{et al.} \cite{LiPRL} calculated a characteristic FEF (denoted here by $\gamma_\text{Zh}$) by considering the real-charge distribution of a very-long, capped single-walled CNT (SWCNT). Their FEF was defined as the ratio $\langle F_{\ell}\rangle / F_\text{M}$ between an average field, $\langle F_{\ell}\rangle$, calculated over a range of about $0.35$ nm above the CNT apex, and $F_\text{M}$. They concluded that $\gamma_\text{Zh}$ would depend on $F_\text{M}$.

In later theoretical work \cite{JAPForbes}, linear dependence of an extracted characteristic FEF on $F_\text{M}$ was found for a capped SWCNT. In this case, a slightly modified definition was used, in that the characteristic FEF was calculated at a fixed position in space, close to the position where the ``local real field" had its maximum absolute value.

However, these theoretical predictions are NOT compatible with the experimental observation that there are CNT current-voltage data sets that lead to linear FN plots that pass the orthodoxy test and yield a constant value for characteristic FEF $\gamma_{\text{C}}$.

Thus, at this point in the argument, which is the situation that existed a couple of years ago (say in 2017), CNT FE theory was in a state comparable with the state of metal FE theory in the early months of 1928, after Lauritsen \cite{LaurPhD,MilLaur28} had discovered the linearity of a plot of the form ``current" vs 1/``voltage", and before Oppenheimer \cite{Opp28} and Fowler and Nordheim \cite{FowlerN} had been able to explain this linearity (albeit in slightly different theoretical ways).

However, there is an instructive difference between the metal FE and the CNT FE situations. In the metal FE case it took less than a year to explain the observed experimental linearity. But, in the CNT FE case, by 2017 it had been over 15 years (since the Dean-Chalamala \cite{Dean2000} experiments) without a satisfactory quantum-theoretical explanation of the observed experimental linearity. Obviously, it has been possible in practice to analyze CNT FE results by making use of classical-conductor emitter models (most commonly the ``hemisphere-on-cylindrical-post" (HCP) model), but the quantum-theoretical justification for doing so was lacking.

Our previous paper \cite{JPCC2019}, called here ``C19", showed the following. After a density-functional-theory (DFT) calculation \cite{parr1994density} of the real-charge-density distribution, we evaluated the related induced-charge-density distributions. We then evaluated the related real-field and induced-field distributions, and the corresponding distributions of real-FEFs and induced-FEFs. For simplicity in a first discussion, we evaluated these things on the symmetry axis of a carbon nanostructure. What we found was that the real-FEFs did depend on the value of macroscopic-field $F_\text{M}$ but the induced-FEFs did not.

This finding, about the apparent importance of considering the induced-charge-density distribution, is broadly compatible with older work by one of us (e.g., \cite{Forbes79}), in the context of the electrostatics of charged metal surfaces as this affects field ion emission.

However, for the matters under discussion, a possible weakness in C19 was that the nanostructure considered was a so-called ``VAGCNT", which is a vertically aligned, CNT/graphene, hybrid three-dimensional system considered to have potential applications in some FE contexts \cite{Talapatra}. It is possible to argue that--particularly when the height of the nanoprotrusion is relatively small, as it is in our case--a VAGCNT may not accurately represent the physical properties of the more common ``post-on-plane" CNT configuration, particularly in respect of image effects in the plane. Thus, in the present paper we carry out, for floating CNTs, procedures analogous to those we applied to VAGCNTs in C19. We address these questions from the perspective of first-principles electronic-structure methods, using density-functional theory (DFT) calculations in the form of the SIESTA code \cite{Soler2002}.

\begin{figure}[h!]
\includegraphics [width=8.8cm,height=4.8cm] {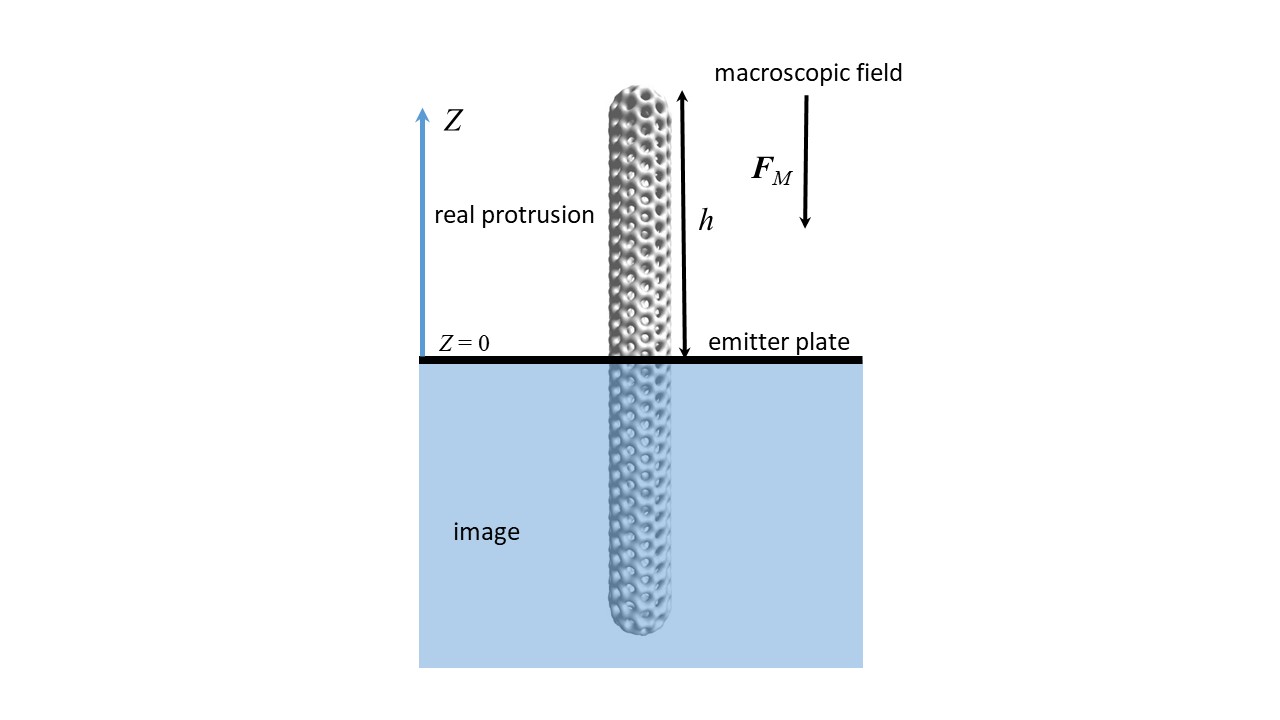}
\caption{Illustration of a field electron emission system involving a (6,6) SWCNT. The classical equivalent is an axially symmetric cylinder with radius $\varrho$, capped at both ends with hemispheres of the same radii $\varrho$. This system is compatible with the classical electron thermodynamics/electrostatics of the usual post-on-plane HCP model, where the post has height $h$.}
\label{floatingCNT}
\end{figure}

Our work focuses on the ``basic electrostatics" of CNT FE, on the grounds that issues of this kind (with their implications for defining the form of tunneling barriers) need to be resolved before emission-physics aspects can be addressed in detail.

In the electrostatics, two fundamental questions arise. (A) How and at which point should the characteristic FEF be calculated for a CNT? (B) Is this calculated FEF a field-dependent quantity or is any such discovered dependence a consequence of using the real-charge distribution to evaluate it?

Question (A) is relevant in the context of classical-conductor models of CNT FE, especially the HCP model, since (for ideal emitters) such models appear to be able provide physically satisfactory explanations of how characteristic-FEF values extracted from FN plots can be related to the physical dimensions (typically radius $\varrho$ and total-height $h$) of the CNT. There is a need to fit the HCP model to the actual atomic-level CNT geometry.

In particular, there is a need to define where, on the CNT symmetry axis ($z$-axis), and relative to the $z$-coordinates of the topmost carbon-atoms, the HCP-model apex (which serves as the ``characteristic location 'C' " used in conventional FE theory) is to be placed. In classical-conductor models of flat atomically-structured metal surfaces, it is known \cite{ForbesES} that the conductor surface should be co-located with the real metal's $\textit{electrical surface}$, which is outside the plane of the metal surface-atom nuclei by a $\textit{repulsion distance}$ equal to about half the nearest-neighbor spacing in the crystal lattice. This result is broadly compatible with more recent work \cite{JAP2016DFT}, using DFT methods , on a pyramidical tungsten emitter.

For CNTs, because of the possibility of field penetration and/or zero-applied-field charge transfers, the situation is much more complicated. For simplicity in making comparisons in the present work, we place the HCP-model apex at the average $z$-coordinate of the topmost carbon-atom nuclei, and take the post radius as equal to CNT radius as defined by the (average) radial position of the carbon-atom-nuclei in the cylindrical section of the CNT. (But we expect that in more precise work, at a later stage, small adjustments in position and radius will be made.)

\begin{figure}[h!]
\includegraphics [width=7.8cm,height=5.8cm] {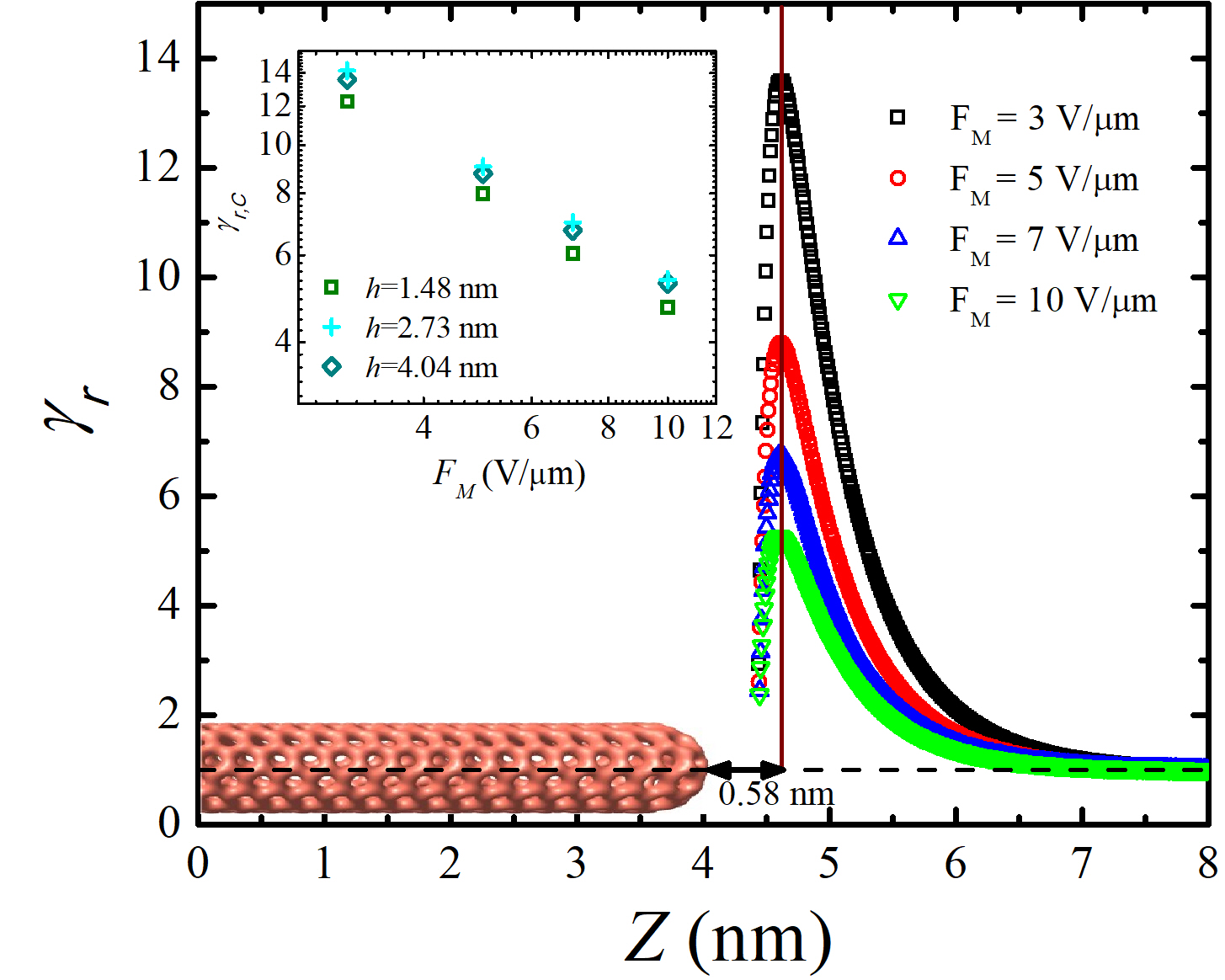}
\caption{$\gamma_\text{r}$, calculated along the the symmetry axis of the floating (6,6) CNT with $2h \approx 8.08$nm, as a function of $Z$ for various values of $F_\text{M}$. An isosurface of $\rho_{\text{r}}(\textbf{\textit{r}}, F_\text{M})$, for $F_\text{M}=10$ V/$\mu$m  is shown at the left inset, with its symmetry axis represented by a dashed line. The vertical (solid) line contains the points where $\gamma_\text{r}$ is maximum, i.e., about $0.58$ nm above the CNT apex. The inset shows shows the maximum value of the real-FEF $\gamma_\text{r,C}\equiv\max\left\{ \gamma_\text{r} \right\}$ as a function of $F_\text{M}$ for the three protruding structures studied in this work, using a log-log scale.} \label{rhoreal}
\end{figure}

We now move on to examine questions (A) and (B) in more detail. We shall use small-length floating-SWCNT models, as illustrated in Fig. \ref{floatingCNT}, with the SWCNT capped at both ends, and with a macroscopic field $F_\text{M}$  externally applied. As noted earlier, a floating SWCNT of length $2h$ that is basically symmetrical about its mid-plane is equivalent  (in terms of its electrostatics and electron thermodynamics) to the usual ``post-on-a-conducting plane" configuration, when the post height is $h$. But the DFT analysis of the floating SWCNT is much easier.

\begin{figure}[h!]
\includegraphics [width=8.0cm,height=5.5cm] {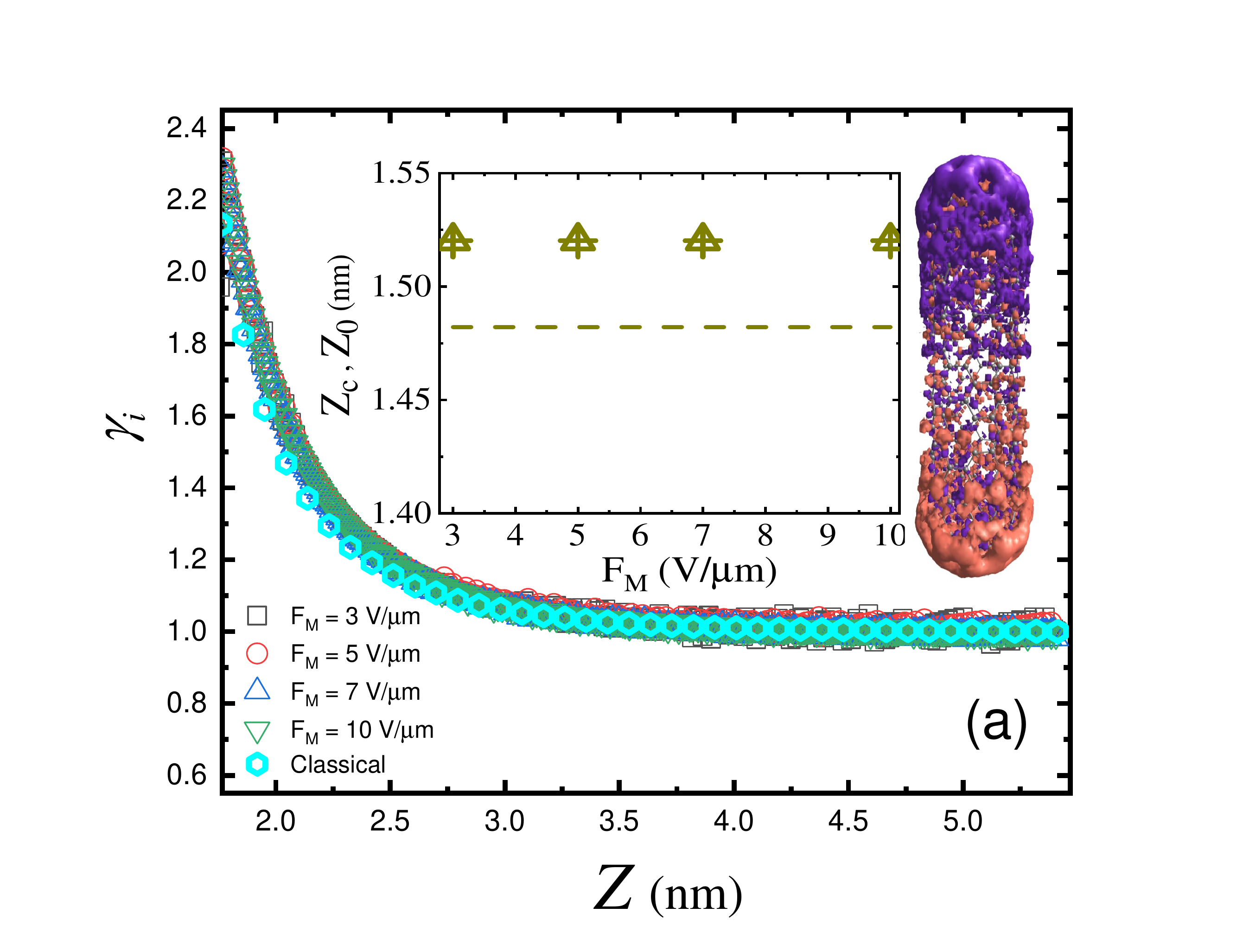}
\includegraphics [width=8.0cm,height=5.5cm] {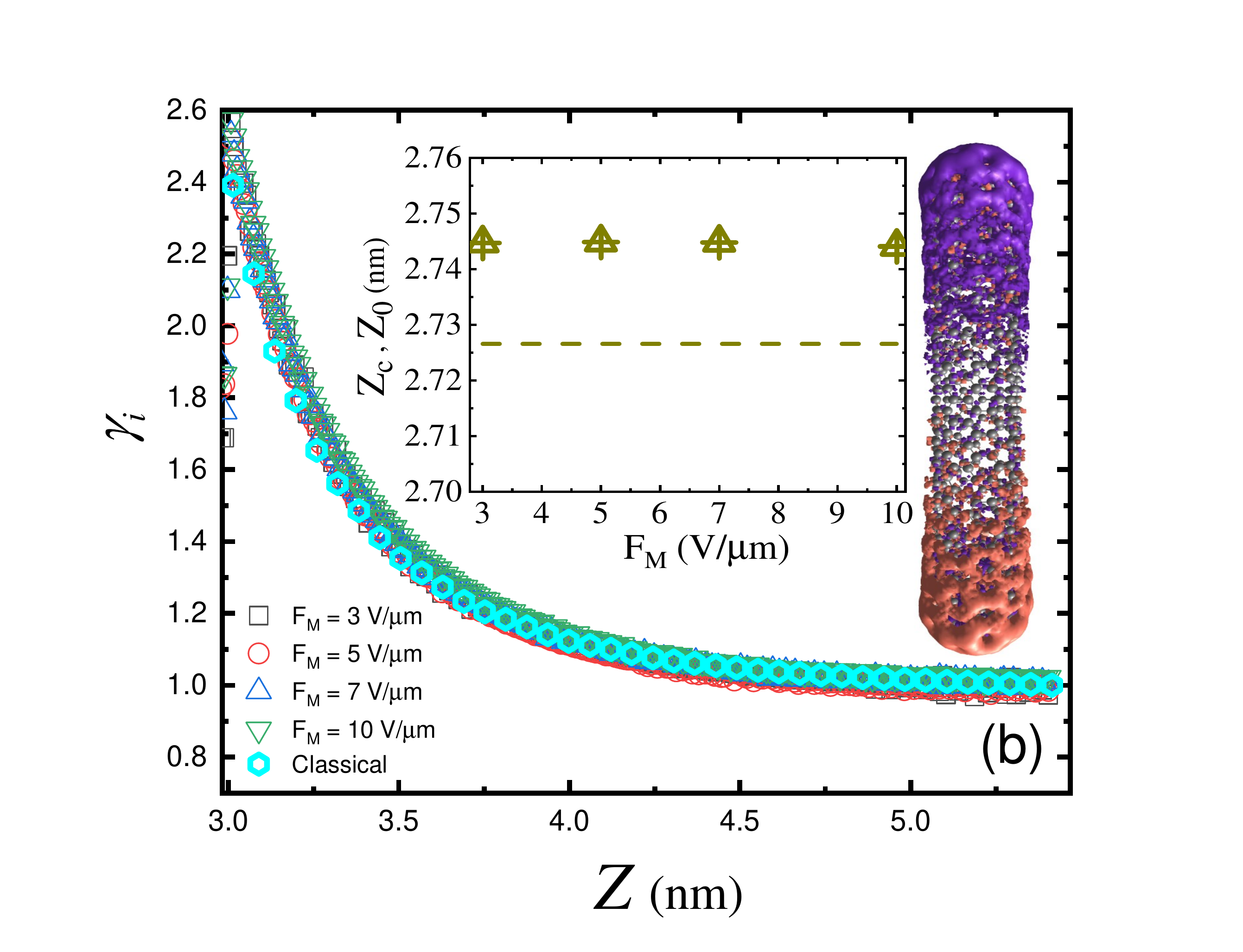}
\includegraphics [width=7.85cm,height=5.5cm] {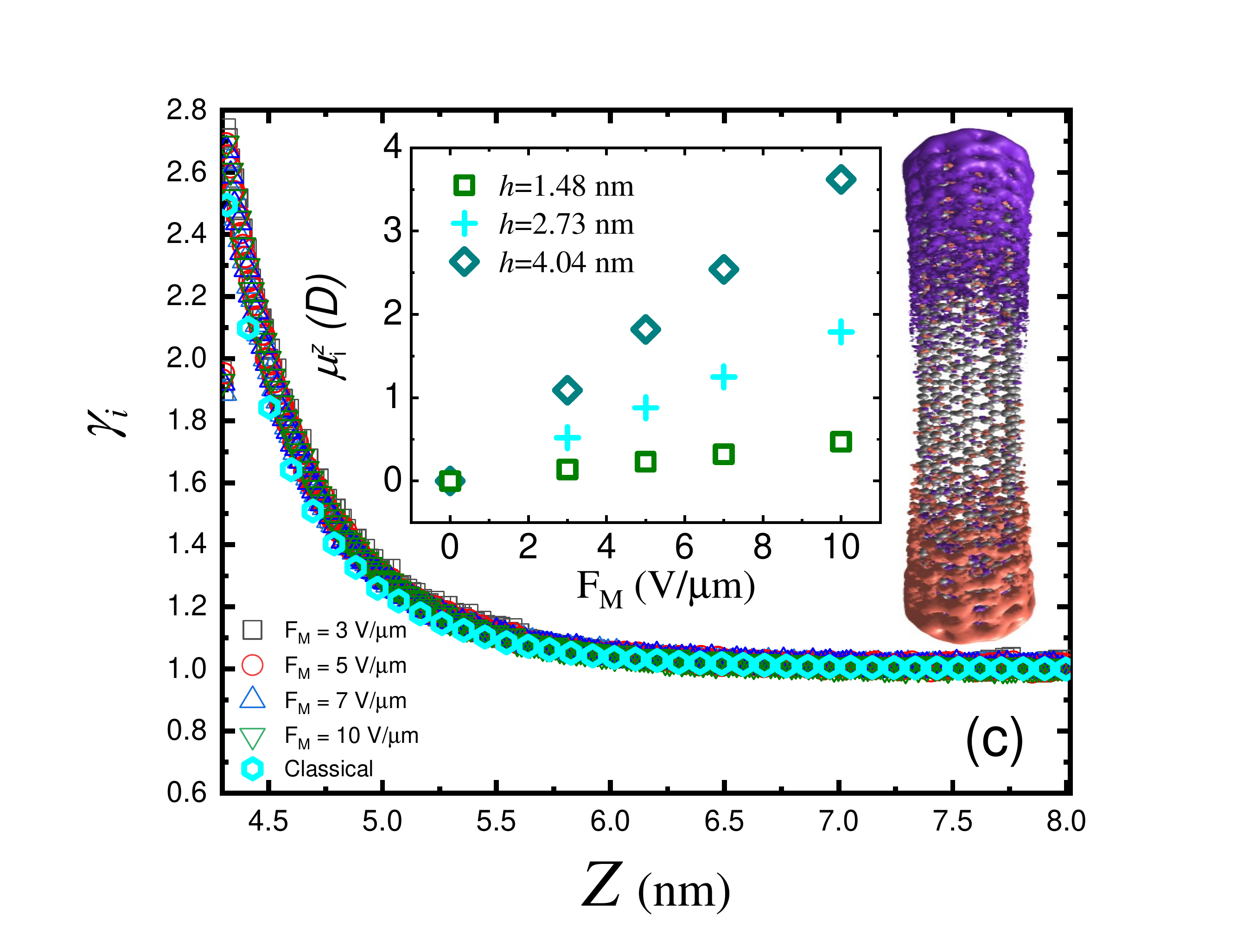}
\caption{Local induced FEFs (LIFEFs) as a function of $Z$, for floating CNTs with: (a) $2h \approx 2.96$ nm; (b) $2h \approx 5.46$ nm; and (c) $2h \approx 8.08$ nm. Results for the HCP classical-conductor model are also shown. The illustrations at the right-hand side are maps of the induced-charge-density distributions for macroscopic field $F_\text{M}$=10 V/$\mu$m (purple=negative, red=positive), positive $Z$ upwards); these show that the CNTs respond by developing a dipole moment, $\mathbf{\mu^{Z}_i}$, along the $Z$-direction. The insets of (a) and (b) show the average positions $Z_0$ of the top-ring atoms (dashed lines), and the positions $Z_c$ of the centroid of the induced charge (symbols), as a function of $F_\text{M}$. The inset of (c) shows the values of $\mu^{Z}_i$, as a function of $F_\text{M}$, for the three CNTs. A very clear linear regime is observed.} \label{HCPBC2}
\end{figure}

We have considered three floating capped (6,6) SWCNTs, with total heights $2h \approx 2.96$, $5.46$ and $8.08$ nm. All these structures have a subnanometer radii $\varrho\approx 0.42$ nm. The structures of the systems were fully optimized within DFT as implemented in the SIESTA code. Energy convergence criteria and box sizes are reported in Ref. \cite{CNTdata}. The electronic structure calculations \cite{RefCalc,PBE} were carried out in the presence of a uniform macroscopic field with magnitude varying between $0$ and $10$ V/$\mu$m.

When making comparisons with HCP-model results, it is important to realize that these  systems in fact behave as arrays of emitters, because using a simulation box of finite size is equivalent to using periodic boundary conditions. The lateral distance between the CNT-model and its nearest images varies from $2.7h$ to $h$, from the smallest to the largest protruding structure. In this range, electrostatic depolarization effects (commonly called ``shielding" or ``screening'') are expected to occur and can be taken into account \cite{Agnol2018}.

Figure \ref{rhoreal} shows values of the local real-FEF (LRFEF), calculated using the ``real" total charge-density distribution (carbon-atom nuclei and electrons), and the applied macroscopic field. These LRFEF values $\gamma_\text{r}$ are values on the CNT symmetry axis and are shown as a function of the coordinate $Z$ that denotes distance measured from the emitter plate as marked in Fig. \ref{floatingCNT}. The results relate to the longest SWCNT ($2h \approx 8.08$ nm), and are shown for various values of $F_\text{M}$.

The issue of deciding how to choose a ``characteristic point" at which a ``charcteristic LRFEF" should be measured is not straightforward. For the time being, we choose the point at which the maximum LRFEF value occurs, on the grounds that this is what is done in a classical-conductor post-like model, where the characteristic FEF is taken as the apex FEF, which is the highest FEF value on the post's symmetry axis.

All three curves (for the different $F_\text{M}$-values) peak at very nearly the same $Z$-value, namely $Z\approx0.58$ nm. However, it is clear that the LRFEF value depends on the macroscopic-field value. This finding is qualitatively similar to that found in the earlier work in ZhiBing Li's group at Sun Yat-Sen University \cite{LiPRL,JAPForbes}, even though they were using much longer CNTs and a (necessarily) less precise quantum-theoretical method. However, we are able to observe a power-law dependence, on $F_\text{M}$, of the maximum on-axis LRFEF, called $\gamma_\text{r,C}$. This is shown in the inset of Fig. \ref{rhoreal}, for all of the three CNT lengths considered here. This result contrasts with the linear relationship found in Ref. \cite{JAPForbes} for capped SWCNTs. We also find that, for a given $F_\text{M}$-value, $\gamma_\text{r,C}$ does NOT increase monotonically with the aspect ratio (i.e., $h/\varrho$) of the half-SWCNTs.

With the floating SWCNTs considered here, the main conclusion at this point is that, if the real-charge-density distribution is used for calculating the FEF of a real CNT, then the results fail to reproduce two well established features of both classical-conductor theoretical results and/or (more important) experimental results from ideal SWCNT emitters. These established features are that the characteristic FEF is independent of macroscopic field, and that the characteristic FEF increases as the SWCNT aspect-ratio increases. Possible failure of the classical HCP model to describe the FEFs of real CNTs might be acceptable in principle, but incompatibility with experiment \cite{Dean2000,BonardPRL2002,BonardPRB} is not.

We now go on to show that if, instead of LRFEFs, the behavior of local induced FEFs (LIFEFs) is considered, then results in agreement with classical-conductor models, and in agreement with experiments, are achieved. As indicated in eq. (1) above, the induced-charge-density distribution is obtained from the real-charge-density distribution by subtracting off the zero-$F_\text{M}$ real-charge-density distribution.

Similarly, we can define a local induced field $F_{\text{i}}$ by
\begin{equation}
F_{\text{i}}(\textbf{\textit{r}}, F_\text{M})\equiv F_{\text{r}}(\textbf{\textit{r}}, F_\text{M}) - F_{\text{r}}(\textbf{\textit{r}}, 0),
\label{indF}
\end{equation}

and then a local induced FEF (LIFEF) by
\begin{equation}
\gamma_{\text{i}}(\textbf{\textit{r}}, F_\text{M}) \equiv F_{\text{i}}(\textbf{\textit{r}}, F_\text{M})  /  F_\text{M}.
\label{LIFEF}
\end{equation}

Figures \ref{HCPBC2} (a), (b) and (c) show $\gamma_\text{i}$ as a function of $Z$ for the three capped SWCNTs studied here.
In each case, results are shown for the $F_\text{M}$-values 3, 5, 7, and 10 V/$\mu$m. These results lie on top of each other, showing that the LIFEF is independent of $F_\text{M}$ at all positions on the symmetry axis that have been examined. It is also found that, for a given SWCNT length, all four curves peak at the same position, and that the LIFEF value at this peak position is that same for all four macroscopic-field values investigated. We also find that in each case the peak position has a $Z$-value that is greater, by about 0.28 nm, than the average position of the six atoms in the topmost SWCNT ring.

In each case, classical electrostatic FEF-values have been calculated by finite-element methods (using COMSOL), after fitting an HCP model to the SWCNT emitter, as described earlier. These results are shown as circles, and closely (but not exactly) overlie the quantum-theoretical LIFEF results.

Between them, these results tentatively suggest that, as in the case of metals, there may exist a relatively well-defined ``electrical reference point", slightly above the CNT apex, that can function as a ``characteristic location" for defining a characteristic LIFEF or a characteristic classical FEF.

For the CNTs, this reference point is further away from the plane of the surface atom nuclei than it is in the case of flat atomically structured surfaces. This would be understandable if it is assumed that the mutual electrostatic depolarization effects that act on the surface atoms of a charged conductor \cite{Forbes79} are weaker for CNTs (because of their pointed shape) than they are for flat metal surfaces. This was also assumed in Ref. \cite{JAPForbes}, where the strongly polarized and dipole-like nature of the atomic charge distribution of the topmost surface atoms, when the macroscopic field is present, is very evident from the quantum-mechanical calculations.

More generally, the most important result of the present work is that, as far as CNT ``electrostatics" is concerned, there seems to be a close link between the behaviour of classical electrostatic models and the behaviour of $\textit{induced-charge-related}$ quantities deduced from quantum-theoretical models. To those scientists who have accepted the validity of this proposition for 30 years or more, when investigating how the atomic-level structure of metals affects the processes that determine the operation of the field ion microscope and related techniques, the idea that a similar argument applies to CNTs seems only natural. However, it does currently seem to be an unfamiliar proposition in the context of CNT physics.

It seems plausible that (in both electric-field polarities) a more basic theoretical formulation/explanation of this result could be provided by static linear response theory. In the CNT context, one considers the external macroscopic field to cause a perturbation in the ground state external potential of the system,  $V_\text{ext}(\textbf{\textit{r}})$. This perturbation leads to a small change in the zero-$F_\text{M}$ electron density, with this small change $\Delta\rho_{\text{r}}(\textbf{\textit{r}}, F_\text{M})$  given by the first term in the expansion
\begin{equation}
\begin{split}
\Delta\rho_{\text{r}}(\textbf{\textit{r}}, F_\text{M}) = \int\chi_\text{e}(\textbf{\textit{r}},\textbf{\textit{r}}')\Delta V_{\text{ext}}(\textbf{\textit{r}}')d\textbf{\textit{r}}' + O[(\Delta V_\text{ext})^2].
\end{split}
\label{response}
\end{equation}
In Eq. (\ref{response}), the function $\chi_\text{e}(\textbf{\textit{r}},\textbf{\textit{r}}')$
can be referred to as the $\textit{static density response function}$. Linear response theory is concerned with its properties and functional form.

As known from this theory, the static density response function is characteristic of the particular system concerned, and is a functional of the ground state density \cite{parr1994density}. In this sense, if a characteristic induced-FEF is defined in terms of the induced electron density, then it will be to some extent dependent on the response function $\chi_\text{e}(\textbf{\textit{r}},\textbf{\textit{r}}')$. Thus, an induced-FEF as defined here can be seen as a property of the CNT that, for a given geometry, is a consequence of how the real-charge-density distribution responds to the application of the macroscopic field $F_\text{M})$, in a linear regime [see inset of Fig. \ref{HCPBC2} (c)].

It needs stressing that many questions remain to be explored/answered. In particular, these include the role of the ``chemically/geometrically-induced" charged transfers that already exist in zero macroscopic field, and give rise to zero-field dipole effects and effects equivalent to the ``patch-field" effects that occur with metals \cite{Her49}. Dipole effects of this kind were also recognized to exist in the earlier work of ZhiBing Li's group (e.g., \cite{JAPForbes}). But issues also include: (a) questions as to whether the same effects would exist if the SWCNT were very much longer, or if (with models of the present length) the macroscopic field were very much larger; and (b) questions about what would be observed if plots were made along a line through the nucleus of one of the topmost carbon atoms, rather than along the symmetry axis. (Preliminary explorations, where possible, suggest that results would be essentially similar.) There is also the unexplained issue of why it is that the induced-FEF seems to compare better with the experimental results, when common sense apparently suggests that FE ought to be determined by the real-potential distribution.

The simplifications used in the present treatment were intended as a way in to discover both basic results and what issues need to be explored more carefully. To summarize, our main result is that, with floating SWCNTs, there is a better match between experiments, classical-conductor models and ``induced-charge-distribution parameters" than there is if the ``real-charge-distribution" parameters are used.


\begin{thebibliography}{29}
\bibitem [{CNT()}]{CNTdata}%
  \BibitemOpen
  \href@noop {} {\bibinfo  {journal} {All carbon atoms are allowed to relax
  until atomic forces decrease approximately below $0.5$ eV/nm. The two
  smallest structures were built in rectangular boxes with sizes $4 \times 4
  \times 11$ nm$^{3}$, whereas the largest one was defined in a box with sizes
  $4 \times 4 \times 16$ nm$^{3}$}\ }\BibitemShut {NoStop}%
\bibitem [{Ref()}]{RefCalc}%
  \BibitemOpen
\bibfield  {journal} {  }\href@noop {} {\bibinfo  {journal} {To solve the
  Kohn-Sham equations, we adopt the PBE exchange-correlation potential. Core
  electrons were described in terms of norm-conserving pseudopotentials and
  valence electrons with double-$\zeta$ basis sets, including polarization
  functions, using an energy cutoff of $300$ Ry and sampling the $\Gamma$ point
  of the Brillouin-zone}\ }\BibitemShut {NoStop}%
\end{thebibliography}

%


\end{document}